\documentclass[aps,twocolumn,groupedaddress]{revtex4}
\usepackage{graphicx}
\usepackage{amsmath}
\usepackage{amssymb}
\begin{document}

\title{On inferring the ion selectivity of the KcsA potassium channel using the distribution of coordination states of the aqueous ions}
\author{P. D. Dixit}
\author{S. Merchant}
\author{D. Asthagiri}\thanks{Corresponding author: Fax: +1-410-516-5510; Email: dilipa@jhu.edu}
\affiliation{Department of Chemical and Biomolecular Engineering,  Johns Hopkins University, Baltimore, MD 21218}

\date{\today}
\begin{abstract}
The S$_2$ site of the KcsA K$^+$ channel has eight carbonyl ligands in the
ion-binding site. A recent study suggests that the K$^+$-over-Na$^+$ selectivity of the 
S$_2$ site can be understood by noting the larger free energy change involved in 
enforcing an eight water  coordination state around Na$^+$ relative to K$^+$.  The free energies
were obtained from the probabilities of observing eight water molecules within  
a coordination sphere whose radius ($\lambda_{\rm Na^+} \approx 3.1$~{\AA} and $\lambda_{\rm K^+} \approx 3.5$~{\AA}) extends to the first minimum of the ion-water oxygen pair correlation function. Curiously, using the same coordination radius led to results that question the
very idea of using coordination states in water to understand selectivity in the channel.  We show that density fluctuations in neat water at the length scale of the coordination volume enter the description of the hydration thermodynamics of the ion expressed in terms of its coordination states. Density
fluctuations explain the sensitivity to the choice of radius. After accounting for 
this effect, the results lead to the conclusion that free energy changes involved in the transition
between coordination states in water are inadequate to explain selectivity in the ion-channel.  
\end{abstract}

\maketitle
\section{Introduction}
Molecular simulations of the selectivity of the S$_2$ site in the KcsA K$^+$ ion 
channel show that \cite{Noskov:2004p77,roux:jgp07,dixitpd:bj09}
\begin{eqnarray}
\Delta \mu^{\rm ex} &=& [\mu^{\rm ex}_{\rm Na^{+}}(\rm S_2) - \mu^{\rm ex}_{\rm K^{+}}(\rm S_2)] - [\mu^{\rm ex}_{\rm Na^{+}}(\rm aq) - \mu^{\rm ex}_{\rm K^{+}}(\rm aq)] \nonumber \\
& \equiv & \Delta \mu^{\rm ex}(\rm S_2) - \Delta \mu^{\rm ex}(\rm aq) \nonumber \\
& \approx & 6~{\rm kcal/mol}. 
\label{eq:mu_basic}
\end{eqnarray}
Here $\Delta\mu^{\rm ex}$ is the free energy of transferring Na$^+$ from the bulk to the
S$_2$ site relative to the same quantity for K$^+$.  $\mu^{\rm ex}$ is 
 the excess chemical potential of the ion; $\mu_{\rm X}^{\rm ex}$(aq) ($\rm X = K^+, Na^+$) is the  hydration free energy and  $\mu^{\rm ex}_{\rm X}(\rm S)$ is the analogous quantity in the S$_2$ site. 

\subsection{Inferring selectivity from the distribution of coordination states of the aqua-ions}
In an effort to rationalize the selectivity of the channel,  Bostick and Brooks \cite{brooks:pnas07} suggest that the selectivity of the S$_2$ site can be understood based on the probabilities of observing specified $n$-coordinate structures ($\rm X[H_2O]_n$) of the aqua-ion in bulk water. (The identification of a particular $n$-cluster requires the specification of a coordination sphere around the aqua-ion.) The idea is that the lower probability of observing the eight-coordinate state of Na$^+$(aq) 
relative to K$^+$(aq) indicates a higher free energy change involved in enforcing 
an eight coordinate state around Na$^+$(aq) relative to K$^+$(aq). But the S$_2$ site also provides
an eight-ligand site, albeit the ligand is a carbonyl and not water.  But the behavior
in water itself suggests that  the ``control of the  permeant ion's coordination state" \cite{brooks:pnas07} is the basis of selectivity.

\begin{figure}[h!]
\includegraphics[scale=0.8]{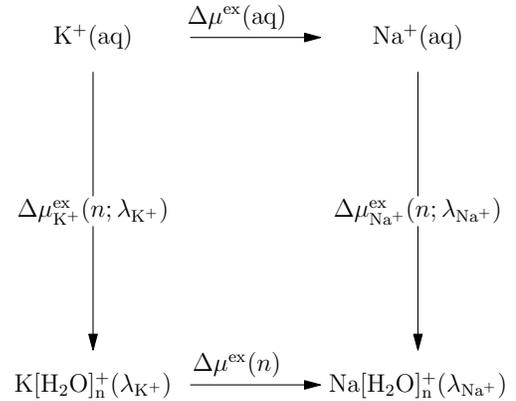}
\caption{Thermodynamic cycle for calculating $\Delta\mu^{\rm ex}(n) - \Delta\mu^{\rm ex}({\rm aq})$
 in terms of $\Delta\mu^{\rm ex}_{\rm K^+}(n,\lambda_{\rm K^+})$ and $\Delta\mu^{\rm ex}_{\rm Na^+}(n,\lambda_{\rm Na^+})$, where $\Delta\mu^{\rm ex}_{\rm X}(n,\lambda_{\rm X}) = \mu^{\rm ex}_{\rm X}(n,\lambda_{\rm X}) - \mu^{\rm ex}_{\rm X}({\rm aq})$. The hydration free energy of the ion X in the coordination
 state $n$ (within a coordination sphere of radius $\lambda_{\rm X}$) is $\mu^{\rm ex}_{\rm X}(n,\lambda_{\rm X})$. The probability of observing the $n$-coordinate state $\rm X[H_2O]_n$ within the coordination sphere is $x_{\rm X}(n,\lambda_{\rm X})$. } \label{fg:cycle}
\end{figure}

Bostick and Brooks \cite{brooks:pnas07} seek $\Delta\mu^{\rm ex} = \Delta\mu^{\rm ex}(n) - \Delta\mu^{\rm ex}({\rm aq})$ (Fig.~\ref{fg:cycle}), and they express this in terms of the free energy change in restricting the coordination state of the aqua-ion to $n$. Thus 
\begin{eqnarray}
\Delta\mu^{\rm ex} & = & \Delta\mu^{\rm ex}(n) - \Delta\mu^{\rm ex}({\rm aq}) \nonumber \\
                                  & = & \Delta\mu^{\rm ex}_{\rm Na^+}(n,\lambda_{\rm Na^+}) - \Delta\mu^{\rm ex}_{\rm K^+}(n,\lambda_{\rm K^+}) \; ,
\label{eq:basic}
\end{eqnarray}
where $\lambda_{\rm X}$ is the coordination radius of ion X and $n$ is the coordination number under consideration. ($n=8$ is used for inferring the selectivity of the channel \cite{brooks:pnas07}.)
Eq.~\ref{eq:basic} is a thermodynamic truism. 

It is in relating $ \Delta\mu^{\rm ex}_{\rm X}(n,\lambda_{\rm X})$ to the probability of observing
the $n$-coordinate state, $x_{\rm X}(n)$, that the earlier analysis is deficient. 
In particular, 
\begin{eqnarray}
\Delta\mu^{\rm ex}({\rm aq}\rightarrow n)^\prime  =  -k_{\rm B}T \ln \frac{x_{\rm Na^+}(n,\lambda_{\rm Na^+})}{x_{\rm K^+}(n,\lambda_{\rm K^+})} 
\label{eq:wrong0}
\end{eqnarray}
is erroneously equated to $\Delta\mu^{\rm ex}$ of Eq.~\ref{eq:basic} (cf. Eq.~10 in Supplementary Information to Ref.~\cite{brooks:pnas07}); the prime on $\Delta\mu^{\rm ex}({\rm aq}\rightarrow n)^\prime$ is to note this error. Eq.~\ref{eq:wrong0} implies that (to within an additivity constant that is the same for both K$^+$ and Na$^+$ and without loss of generality taken equal to zero),
\begin{eqnarray}
\Delta\mu^{\rm ex}_{\rm X}(n,\lambda_{\rm X}) = -k_{\rm B} T \ln x_{\rm X}(n,\lambda_{\rm X}) \; .
\label{eq:wrong1}
\end{eqnarray} 
It is rather surprising that the right hand side of Eq.~\ref{eq:wrong1} is always positive but there is no obvious physical reason that the left hand side of Eq.~\ref{eq:wrong1} must always be positive. 
Further since $\sum_n x_{\rm X}(n,\lambda_{\rm X}) = 1$, Eq.~\ref{eq:wrong1}  leads to 
\begin{eqnarray}
 e^{-\beta \mu^{\rm ex}_{\rm X}}  =  \sum_{n} e^{-\beta \mu^{\rm ex}_{\rm X}(n,\lambda_{\rm X})} \label{eq:muInv_incorr} \; .
\end{eqnarray}

If instead of the transition between the aqua-ion to the $n$-coordinate state of the ion, 
the transition between the most probable coordination state of the ion ($\tilde{n}$) to the 
$n$-coordinate state is considered, Eq.~\ref{eq:wrong0} becomes
\begin{eqnarray}
\Delta\mu^{\rm ex}(\tilde{n}\rightarrow n)^\prime  & = &  -k_{\rm B}T \ln \frac{x_{\rm Na^+}(n,\lambda_{\rm Na^+})}{x_{\rm Na^+}(\tilde{n},\lambda_{\rm Na^+})} \nonumber \\
& + & k_{\rm B} T \ln \frac{x_{\rm K^+}(n,\lambda_{\rm K^+})}{x_{\rm K^+}(\tilde{n},\lambda_{\rm K^+})}  ,
\label{eq:wrong2}
\end{eqnarray}
where different values of $\tilde{n}$ are implied for Na$^+$ and K$^+$. In the earlier study (Fig.~3B, Ref.~\cite{brooks:pnas07})  the ``most popular coordinated states"\cite{brooks:pnas07} were indicated
on the free energy landscape (at a free energy  value near zero) on which the $n=8$ coordinate case was also noted. Based on this, earlier we \cite{dixitpd:bj09} had concluded that Bostick and Brooks \cite{brooks:pnas07} had used Eq.~\ref{eq:wrong2}, when in fact they were using Eq.~\ref{eq:wrong0}.

\begin{figure}
\includegraphics[scale=0.9]{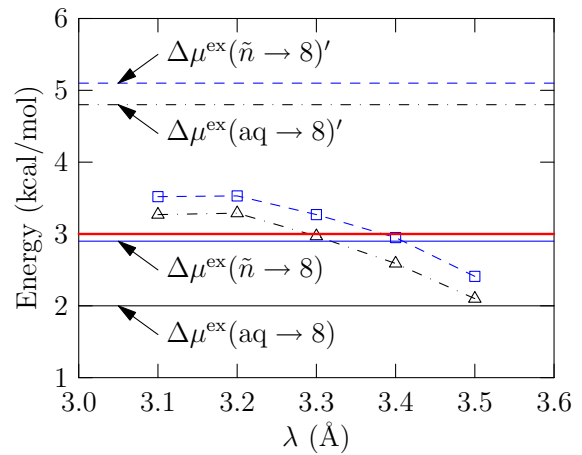}
\caption{Estimates of the selectivity with various approaches. {\em Red line\/}: the
stated lower bound of selectivity \cite{brooks:pnas07}. Theories that lead to an estimate below this line would be regarded as inadequate in explaining selectivity. {\em Broken blue and black lines\/}: 
$\Delta\mu^{\rm ex}({\rm aq}\rightarrow 8)^\prime$ is based on Eq.~\ref{eq:wrong0} and $\Delta\mu^{\rm ex}(\tilde{n}\rightarrow 8)^\prime$ is based on Eq.~\ref{eq:wrong2}.  These two estimates differ by about 0.3~kcal/mol ($\approx 0.5\; k_{\rm B} T$). $\lambda_{\rm Na^+} = 3.1$~{\AA} and $\lambda_{\rm K^+} = 3.5$~{\AA}. {\em Unbroken blue and black lines\/}: Calculations using Eq.~\ref{eq:correct0} and
the corrected version of Eq.~\ref{eq:wrong2}. 
{\em Symbols}: Calculations using $\lambda = \lambda_{\rm Na^+} = \lambda_{\rm K^+}$ and the correct forms of Eq.~\ref{eq:wrong0} ($\triangle$) and Eq.~\ref{eq:wrong2} ($\Box$). Simulation methods are
noted in Appendix A.}\label{fg:data} 
\end{figure}

{\bf Results using Eqs.~\ref{eq:wrong0} and~\ref{eq:wrong2}}: Fig.~\ref{fg:data} collects the results for estimates of the selectivity using Eqs.~\ref{eq:wrong0}~and~\ref{eq:wrong2} and the corrected versions of these equations (see below).  Eqs.~\ref{eq:wrong0} and~\ref{eq:wrong2} lead to an estimate of selectivity about 2~kcal/mol higher than the lower bound of selectivity ($\approx 3$~kcal/mol \cite{brooks:pnas07}).  Observe that Eq.~\ref{eq:wrong2} leads to an estimate not much different from that based on Eq.~\ref{eq:wrong0}. This is consistent with the observation \cite{safir:jcp09} that 
in the description of the hydration free energy of the ion in terms of its various coordination states 
(and a correction for long-range interactions), only coordination states with $n\leq \tilde{n}$ contribute dominantly to the thermodynamics. Since for Na$^+$, $\tilde{n} = 6$ ($\lambda_{\rm Na^+} = 3.1$~{\AA}), and for K$^+$, $\tilde{n} = 7$ ($\lambda_{\rm K^+} = 3.5$~{\AA}), it is then reasonable to expect that Eqs.~\ref{eq:wrong0} and~\ref{eq:wrong2} give similar estimates. 
%
%

\section{Density fluctuations and $\Delta\mu^{\rm ex}_{\rm X}(n,\lambda_{\rm X})$}
We first note the following identities that readily follow from a multi-state generalization of the potential distribution theorem \cite{widom:jpc82,lrp:mulGjacs97,lrp:apc02,lrp:book,lrp:cpms,dixitpd:bj09,safir:jcp09}: 
\begin{subequations}
\label{eq:allmulti}
\begin{eqnarray}
 e^{\beta \mu^{\rm ex}_{\rm X}({\rm aq})} & = & \sum_{n} x_X(n,\lambda_{\rm X})  \cdot e^{\beta \mu^{\rm ex}_{\rm X}(n,\lambda_{\rm X})} \label{eq:muInv} \\
  e^{-\beta \mu^{\rm ex}_{\rm X}({\rm aq})} & = & \sum_{n} p(n,\lambda_{\rm X})  \cdot e^{-\beta \mu^{\rm ex}_{\rm X}(n,\lambda_{\rm X})} \label{eq:muFor} \; .
\end{eqnarray}
\end{subequations}
Here $p(n,\lambda_{\rm X})$ is the probability of observing $n$ water molecules in an
ion-free coordination sphere of radius $\lambda_{\rm X}$ in neat liquid water. (We will call the
ion-free coordination sphere the observation volume.) The set $\{p(n,\lambda_{\rm X})\}$ codifies
density fluctuations in neat water at the length scale of the observation volume \cite{lrp:jpcb98};
$p(0,\lambda_{\rm X})$ is an essential factor in the theory of hydrophobic effects  and is a measure of the packing interactions involved in hydration \cite{lrp:apc02,lrp:book,lrp:cpms,lrp:jpcb98}.

\begin{figure*}
\includegraphics{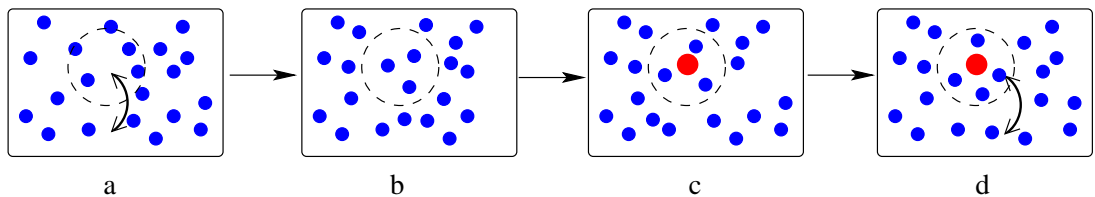}
\caption{Schematic description of Eq.~\ref{eq:xpratio}. The large filled circle is the solute and the small
filled circles represent water molecules. The dashed circle around the solute denotes the coordination sphere and that in neat water is the  observation volume. (a) Neat water with an observation volume that
is open to the exchange of water molecules with the bulk. The arc with arrow-heads on both ends represents the
exchangeability of water.
(b) $p(3)$ is the probability of observing  three water  molecules in the 
observation volume in the ensemble of states (a). The free energy change in going from (a) to (b) is
$-k_{\rm B} T \ln p(3)$. (c) The particle is inserted into the observation volume with 3 water molecules. 
The excess chemical potential of the ion in the 3-coordinate state is $\mu^{\rm ex}(3)$.
(d) The constraint on the number of water molecules in the coordination sphere is relaxed. The probability of state (c) in the ensemble of states (d) is
$x(3)$ and the free energy change in going from (c) to (d) is $k_{\rm B} T \ln x(3)$.  The excess
chemical potential of the ion in (d) is $\mu^{\rm ex}$. Thus $\mu^{\rm ex} = \mu^{\rm ex}(n) + k_{\rm B} T \ln x(3) -k_{\rm B} T \ln p(3)$.}\label{fg:qc}
\end{figure*}

The following relation was derived earlier \cite{safir:jcp09}
\begin{eqnarray}
 e^{\beta \mu^{\rm ex}_{\rm X}({\rm aq})} & = &\frac{x_X(n,\lambda_{\rm X})}{p(n,\lambda_{\rm X}) }  e^{\beta \mu^{\rm ex}_{\rm X}(n,\lambda_{\rm X})} \; . 
\label{eq:xpratio}
 \end{eqnarray}
It can be seen that Eq.~\ref{eq:xpratio} satisfies both members of Eqs.~\ref{eq:allmulti}
and also has an intuitively appealing explanation (Fig.~\ref{fg:qc}). 
%
%
%
%

Using Eq.~\ref{eq:xpratio} in Eq.~\ref{eq:basic}, we obtain 
\begin{eqnarray}
\Delta\mu^{\rm ex} & =  &  -k_{\rm B}T \ln \frac{x_{\rm Na^+}(n,\lambda_{\rm Na^+})}{x_{\rm K^+}(n,\lambda_{\rm K^+})} \nonumber \\
& & +  k_{\rm B}T \ln \frac{p(n,\lambda_{\rm Na^+})}{p(n,\lambda_{\rm K^+})} .
\label{eq:correct0}
\end{eqnarray}
Similarly, an equation for $\Delta\mu^{\rm ex}(\tilde{n}\rightarrow n)$ can be obtained.

Comparing Eq.~\ref{eq:wrong0} to Eq.~\ref{eq:correct0} (or Eqs.~\ref{eq:wrong1} and~\ref{eq:muInv_incorr} to Eqs.~\ref{eq:xpratio} and~\ref{eq:muFor}, respectively), 
 it is evident that  the factor $p(n,\lambda_{\rm X})$ is omitted in the earlier development \cite{brooks:pnas07}. This omission leaves out  physical phenomena of importance 
 in the thermodynamics of hydration, the consequences of which are apparent in Fig.~\ref{fg:data} (the unprimed $\Delta\mu^{\rm ex}$ values).  Including
the role of density fluctuations in neat water at  the length scale of the observation volume 
predicts a selectivity less than the suggested lower bound of selectivity. Thus 
control of the coordination state of the ion alone is an inadequate
explanation of the selectivity of the ion channel.   As suggested earlier \cite{Noskov:2004p77,roux:jgp07,dixitpd:bj09}, acknowledging the chemical difference between the coordinating ligand, water in the aqueous phase and carbonyl in the S$_2$ site,  is necessary to
understand the selectivity of the ion-channel.

When $\lambda = \lambda_{\rm Na^+} = \lambda_{\rm K^+}$, the second term on the right hand side of Eq.~\ref{eq:correct0} is zero. In this instance  Eq.~\ref{eq:correct0} is the same as Eq.~\ref{eq:wrong0}. 
The estimated $\Delta\mu^{\rm ex}$ for $\lambda = 3.5$~{\AA} is about 2~kcal/mol and is in 
fair agreement with the 1~kcal/mole estimate obtained using 
Monte Carlo simulations \cite{dixitpd:bj09}. Most importantly, both these estimates are significantly different from the approximately 5~kcal/mol obtained using Eq.~\ref{eq:wrong0} with $\lambda_{\rm Na^+} = 3.1$~{\AA} and $\lambda_{\rm K^+} = 3.5$~{\AA} (Fig.~\ref{fg:data}).  Thus we can understand the observation that when Eq.~\ref{eq:wrong0} is used with the same coordination radii, a lower selectivity is predicted. The lower selectivity arises due to the cancellation of the effects of water occupancy in an
ion-free coordination volume of the same size for both Na$^+$ and K$^+$. 

Before concluding this section, we present some thoughts on where the earlier \cite{brooks:pnas07} development may have faltered.  The coordination sphere is an open system in contact with an external bath that provides solvent molecules at a constant chemical potential, but the coordination volume and the number of particles it can hold are nowhere close to the thermodynamic limit. Thus care is required in assigning a thermodynamic potential to the partition function of the
system. Further, the small-system under consideration will be sensitive to fluctuations. Hence
describing the system with different ensembles may not be equivalent. Equivalence of ensembles
is strictly valid only in the thermodynamic limit when relative fluctuations in the 
variables conjugate to those used to describe the ensemble are small. 
Thus, perhaps, using an open system framework to study a system whose population is fixed 
(Cf. Eq.~7 in Supplementary Information to Ref.~\cite{brooks:pnas07}) may explain the
disagreement with results presented here. 

The present development avoids the subtleties noted above.  The theoretical framework used here \cite{lrp:mulGjacs97,lrp:apc02,lrp:book,lrp:cpms,dixitpd:bj09,safir:jcp09} is grounded in 
well-established statistical mechanics, and the results (for example, Fig.~\ref{fg:qc}) 
agree with physical reasoning.

\section{Conclusions}
The probability of observing a given number of water molecules in neat water in an
ion-free coordination volume codifies density fluctuations in the liquid at the scale of the
coordination volume. Including this effect suggests that free energy changes involved
in modifying the coordination state of an ion in water \cite{brooks:pnas07}  is 
inadequate in explaining the ion selectivity in the ion channel.

\appendix
\section{Methods}
$NVT$ simulations of neat water and of Na$^+$ and K$^+$\cite{roux:jcpKNa} in water
 were conducted using the NAMD program \cite{namd}. Liquid water was described using the
 TIP3P \cite{tip32} model. A temperature of 298~K was maintained using a Langevin 
 thermostat. The cubic simulation system comprises 306 water molecules for the pure water simulation; the ion-water system comprises an additional ion. The ion is always held fixed at the center of the simulation cell. The total number density (counting water molecules 
and the ion, if present) is 33.33 nm$^{-3}$. For the ion water system, after over 10~ns of equilibration
data was collected for 16~ns. Configurations were saved every 100~fs for further analysis. 
For coordination radii from 3.1~{\AA} to 3.5~{\AA}, the distribution of water molecules in the
coordination shell of the ion was calculated. An extensively equilibrated box of water molecules
was further equilibrated for 100~ps. Configurations were saved every 250~fs during the course of a production run lasting 1.996~ns.  For each configuration, 
observation volumes of radii between 3.1~{\AA} to 3.5~{\AA} were centered on 2197 
points arranged in a cubic lattice within the simulation cell and the distribution of water molecules
calculated.  

\begin{acknowledgements}
We thank David Bostick and Charles Brooks for constructive discussions.  
\end{acknowledgements}

\end{document}